\documentstyle[12pt,epsf]{article}
\def\be{\begin{equation}}
\def\ee{\end{equation}}

\def\ffrac#1#2{\textstyle{#1\over#2}\displaystyle}
\gdef\journal#1, #2, #3, 1#4#5#6{{\sl #1~}{\bf #2}, #3, 1#4#5#6}

\begin{document}
\baselineskip=18pt
\pagestyle{empty}
\vspace{-1mm}
\begin{flushright}
{cond-mat/9806098\\
JLC-98-2}
\end{flushright}
\vspace{5mm}
\begin{center}
{\large{\bf Renormalisation Group Theory\\
 of Branching Potts Interfaces\\}}
\vspace{15mm}
{\bf John Cardy\\}
\vspace{8mm}
{\em Department of Physics\\
Theoretical Physics\\1 Keble Road\\Oxford OX1 3NP, UK\\}
\vspace{4mm}
{\em \& All Souls College, Oxford\\}
\end{center}
\vspace{5mm}
\begin{abstract} We develop a field-theoretic representation for the configurations of an
interface between two ordered phases of a $q$-state Potts model in two
dimensions, in the solid-on-solid approximation. The model resembles
the field theory of directed percolation and may be analysed using
similar renormalisation group methods. In the one-loop approximation
these reveal a simple mechanism for the emergence of a critical value
$q_c$, such that for $q<q_c$ the interface becomes a fractal with a
vanishing interfacial tension at the critical point, while for $q>q_c$
the interfacial width diverges at a finite value of the tension,
indicating a first-order transition. The value of the Widom exponent for
$q<q_c$ within this approximation is in fair agreement with known exact values. 
Some comments are made on the case of quenched randomness.
We also show that the $q\to-\infty$ limit of our model corresponds to
directed percolation and that the values for the exponents in the
one-loop approximation are in reasonable agreement with accepted values.
\end{abstract}
\newpage

\pagestyle{plain}
\setcounter{page}{1}
\setcounter{equation}{0}
\section{Introduction}
The study of interfaces between ordered phases, besides being of interest for
its own sake, often also provides useful information on the bulk
properties of the system in question, especially close to a bulk
critical point. For example, at a continuous transition, the interfacial 
tension is supposed to vanish as $(T_c-T)^\mu$, where the Widom exponent
$\mu$ is related to the conventional bulk correlation length exponent by the
scaling law $\mu=(D-1)\nu$, where $D$ is the overall dimensionality of
the system \cite{Widom}.

When there are only two coexisting phases, as in a ferromagnetic Ising
model, the structure of such an interface is relatively simple,
especially at low temperatures. An interface in a two-dimensional 
system\footnote{We use coordinates $t$ and $x$ parallel and
perpendicular to the interface, for reasons which will become clear.}
may be generated by imposing suitable conditions on the
Ising-like degrees of freedom $s(t,x)=\pm1$ at the boundary of a finite
but large box $(0\leq t\leq L,|x|\leq L/2)$: for example that $s=1$ on 
$x=L/2$, $(0,x>0)$ and $(L,x>0)$,
while $s=-1$ on $x=-L/2$, $(0,x<0)$ and $(L,x<0)$. At low
temperatures the resultant interface has the form of a \em directed path
\em from $(0,0)$ to $(L,0)$, that is, there are no overhangs. In
addition, at low temperatures, bubbles of the wrong phase in the bulk
regions above and below the interface are suppressed. The approximation
of allowing only such directed paths gives the \em solid-on-solid \em
(SOS) model for the interface. It is extremely easy to analyse, for
example by transfer matrix methods, because the directed path may be
thought of as a discrete time version of a simple random walk in the
$x$-direction, with $t$ playing the role of time. Alternatively, the
partition function may be viewed as the imaginary time version of the
Feynman path integral for a non-relativistic particle. Although it is
strictly valid only at low temperatures, the SOS model does in fact
capture accurately several important aspects of the true critical
behaviour as $T\to T_c-$. In particular it yields the exact value 
$\mu=1$ for the Ising model. It is understood that this success arises
from a cancellation between overhang and bubble contributions in certain
quantities which is probably peculiar to the Ising model. Nevertheless
the ease of analysis suggests that it might be useful to extend it to
other systems.

\begin{figure}
\centerline{
\epsfxsize=4in
\epsfbox{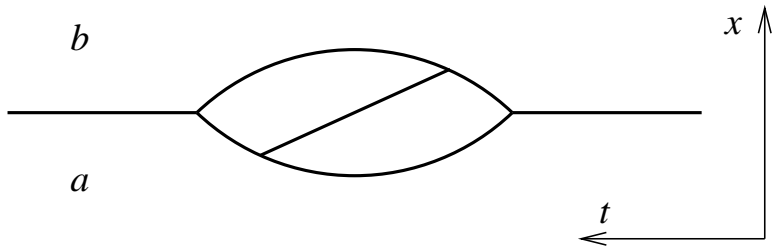}}
\caption{Typical configuration of a branching interface separating ordered
phases $a$ and $b$. This particular set of contributions has weight
$(q-2)(q-3)$.}
\label{fig1}
\end{figure}

When there are more than two coexisting phases, as in general for the
$q$-state Potts model, the structure of the interface is more
complicated, since it can branch as illustrated in Fig.~\ref{fig1}.
At low temperatures, such branching should be suppressed since it
locally increases the interfacial tension, but near the critical point
this can be offset by entropic gains.
Thus it is necessary to take into account all possible branchings
near $T_c$. In general this is very difficult, but once again within
the SOS approximation progress is possible. This is because, in the
random walk analogy, one now has to consider random walks which may
branch and coalesce. Similar processes occur in the field-theoretic
formulation of directed percolation (DP) \cite{CS}, which represents a rather
generic class of non-equilibrium phase transitions. Examples are the
so-called branching and annihilating random walks
(BARW) \cite{BARW} which
have recently been studied rather extensively. Using the Feynman path integral
analogy, or by other methods, it is possible to set up
a field theory decribing the diffusion, branching and annihilation
processes of such many-particle systems. This is then amenable to standard
methods of renormalisation group analysis. 

In fact, branching interfaces within the SOS approximation have 
a number of
important differences from ordinary interacting random walks. 
If Fig.~\ref{fig1} is
simply viewed as a Feynman diagram in configuration space, 
a given propagator joining two vertices
represents a sum over \em all \em directed paths between the vertices,
including paths which may cross those corresponding to other lines in
the diagram. However, it does not make physical sense for interfaces to cross.
This may be taken into account by introducing an infinitely
strong local repulsion
between the particles in the field theory. In one
transverse $x$ dimension, this is equivalent to assuming 
that the particles behave as
fermions, at least in between the branching and coagulation events.
This turns out to be the simplest approach from a calculational point of
view.

The other important difference is in the weighting factors. A given
set of interfacial configurations such as that in Fig.~\ref{fig1} should
be weighted by the number of ways of colouring the diagram with the
$q$ colours of the Potts model, consistent with the requirement that
no two colours on either side of a given segment of interface are the same,
and that the two (different) colours above and below the whole interface
are fixed. A given diagram with $n$ branchings (and $n$ annihilations)
will thus have a weight which is a polynomial in $q$ of degree $n$.
No such weighting factors arise for DP or ordinary BARWs, and it is not
\em a priori \em obvious that they will lead to a diagrammatic structure
which satisfies the correct Schwinger-Dyson equations so as to be 
multiplicatively renormalisable. However, in the next Section we argue
that these diagrams can in fact be derived from a local lagrangian and
therefore should be amenable to conventional renormalisation methods.

One of the well-known features of the bulk $q$-state Potts model is that
there exists a critical value $q_c$ such that, for $q\leq q_c$, the 
transition is continuous, with a diverging correlation length, while
for $q>q_c$ it is first-order. In $D=2$ dimensions $q_c=4$ \cite{Baxter}. 
The physics
behind this change of behaviour is not completely understood. For very
large $q$ it is possible to see from the mapping to the random cluster
model that at the critical point point there are two states (either all
sites are in their own separate cluster, or all sites are in the same infinite
cluster) which have the same free energy but very different internal
energies, corresponding to first-order coexistence. The inclusion of
fluctuation effects does not alter this conclusion to any finite order
in an expansion in powers of $1/q$. Clearly this cannot persist to
$q=2$ (the Ising model) when the transition is continuous. However,
in mean field theory $q_c=2$, and studies within the $\epsilon$-expansion below
six dimensions indicate that this value persists, at least to $D=4$.
On the other hand,
simple real space renormalisation group calculations tend to predict the
transition to be continuous for all finite $q$. It is only when
additional lattice degrees of freedom corresponding to vacancies are
introduced that a scenario is found which allows the emergence of a
non-trivial value of $q_c$ \cite{Nienhuis}. 

From the point of view of the interface, this change in behaviour as a
function of $q$ should be signalled by a vanishing renormalised
interfacial tension at the transition for $q<q_c$, and a finite value
for $q>q_c$. This is indeed what we find 
within the SOS approximation described above. Apart from $q$, the
continuum version of our model contains two parameters: the bare
interfacial tension $\sigma_0$ and the Boltzmann weight $u_0$ for a
branching (or coagulation). On a given lattice these are of
course smooth functions of the reduced coupling $J/kT$ of
the Potts model, but they turn out to renormalise differently. Our
renormalisation group equations are based on the way that the
dimensionless
renormalised branching rate $g$ (which, roughly speaking,
measures the probability of
a large-scale branching of the interface into two separate parts)
varies with the renormalised interfacial tension $\sigma$, keeping the
bare branching weight $u_0$ fixed.  This equation has the form 
\be\label{eq1}
\sigma{\partial g\over\partial\sigma}\equiv\beta(g)=
-\ffrac14g-b(q-q_0)g^3+O(g^5),
\ee
where the right hand side is the result of a one-loop calculation,
with $b$ a positive constant, and $q_0=\ffrac{24}5$.
Within this approximation we already see
interesting behaviour. As $\sigma$ increases (corresponding to $T\to0$),
$g\to0$ as expected. On the other hand, as $\sigma\to0$, for $q<q_0$
we see that $g$ approaches a finite fixed point value, indicating that
at the critical point there is branching on all scales and the interface
is a fractal. For $q>q_0$, on the other hand, integration of (\ref{eq1})
shows that, starting from a finite value for some large low-temperature
value of $\sigma_0$,
$g$ actually diverges at a \em finite \em value of $\sigma$. Although of
course this is outside the region of validity of the one-loop
approximation, it does nevertheless provide a mechanism for the
existence of a lower bound on the renormalised interfacial tension,
indicating that the transition must be first-order. Once $g$ diverges,
the interface will break up into many components and the bulk ordered phases
will no longer be distinct. We may therefore identify this point with
the critical temperature, and $q_0$ with the critical value $q_c$, 
in this approximation.

A more detailed analysis for $q<q_0$ allows the value of the Widom
exponent to be extracted. Within the same approximation, we find
for example $\mu\approx\frac9{11}$ for $q=3$, to be compared with the
exact value of $\frac56$. 

As has been remarked elsewhere \cite{Kardar}, the configurations of the 
interface resemble those of the diagrams in the perturbative expansion for the 
connectedness function in directed percolation (DP). Indeed, it is 
straightforward to show that taking the formal limit $q\to-\infty$ with
$u_0^2|q|$ fixed in our diagrams give precisely the correct weights for DP.
We may therefore compare our one-loop results for the exponents with those
of DP, which are known very accurately in one transverse dimension.
We find $\nu_\parallel\approx1.67$, $z\approx1.6$, and $\eta\approx-0.4$,
in comparison with the accepted values \cite{JG} $1.73$, $1.58$ and $-0.31$ 
respectively.

These are quite close, considering that this is only a simple one-loop estimate.
In fact, this agreement confirms the broad universality of the DP class.
The usual field theory approach to DP involves either an $\epsilon$-expansion
about $d=4$ transverse dimensions,
or a loop expansion of a theory with only cubic couplings and no bare quartic
coupling. Carried to sufficiently high order and resummed, these
give values of the exponents in fair agreement with those of
simulations and enumerations. By contrast, our expansion is in the cubic
couplings at \em infinite \em value of the repulsive quartic coupling.
In principle this could lead to a different universality class. However, the 
good agreement in the exponents suggests that this is not the case and that,
as has been observed elsewhere, DP universality is particularly strong.

We were originally motivated to study this problem in an attempt to
resolve some of the confusion surrounding the behaviour of the $q$-state
Potts model with random bonds. A rigorous result of Aizenman and Wehr
\cite{AW} shows that in this case
the first-order transition should be smoothed for all $q$, 
in the sense that there should be no latent heat. This
suggests, by analogy with other known systems, that 
the interfacial tension should also vanish at the critical point.
Initial Monte Carlo studies for this model for $q=8$ found the
transition indeed to be continuous, with both magnetic and thermal
critical exponents consistent with the Ising values $\beta=\frac18$
and $\nu=\mu=1$ \cite{Chen}. 
However, more recent studies, using both Monte Carlo
methods \cite{Picco}
and finite-size scaling of the transfer matrix \cite{JJ}, have found a
magnetic exponent significantly different from this value, and, moreover,
continuously varying with $q$, although the thermal exponent remains
numerically consistent with $\nu=1$. 

An analysis of branching Potts interfaces on a hierarchical lattice,
in both pure and random models, has been made by Kardar, Stella, Sartoni
and Derrida \cite{Kardar}. For the pure system they were able to write down
exact recursion relations for the interfacial partition function.
However, on this lattice, the transition in the pure system is
continuous for all finite $q$, thus casting doubt on the reliability of
such a model for studying the effect of quenched impurities on
first-order transitions. Nevertheless, this model is sufficiently simple
that the problem is tractable. These authors found that the critical
behaviour of the random interface is controlled by a zero-temperature
RG fixed point at which the branching weight is marginally irrelevant.
This implies that the critical exponents should be independent of $q$.
These authors suggested that a similar mechanism may operate on a
regular two-dimensional lattice, consistent with the original Monte
Carlo results of Chen et al.~\cite{Chen}, but not with the more recent
work \cite{Picco,JJ}.

It would be very interesting to analyse directly the effect of
impurities on Potts interfaces in two dimensions. In studying
this in the SOS
approximation one immediately meets a difficulty. In the bulk Potts
model, quenched bond randomness is marginally irrelevant for $q=2$ and
relevant for $q>2$. The examples of hierarchical lattices in
Ref.~\cite{Kardar} also have this property. However, in the SOS
approximation, the randomness turns out to be strongly relevant even in
the Ising case. Indeed, this is the well-studied problem of a directed
polymer in a random medium, which may also be mapped to the KPZ problem
in one dimension \cite{KPZ}. 
Therefore we conclude that the SOS approximation to
the interface does
not capture the correct physics of the random Ising model, nor, by
extension, that of the Potts model near $q=2$. It is of course, interesting
to enquire what is the effect of branching on directed polymers in a
random medium, even though this has no apparent relevance to the random
Potts model. Curiously, we find that it is again marginal, as in the
models studied by Kardar et al.~\cite{Kardar}. 

There is long history of study of Potts interfaces, begun in the 1980s.
Much of this concerned the wetting properties of the disordered phase
for $q>q_c$
\cite{wetting}, but there were also some attempts to understand the critical
properties for $q<q_c$ through the branching structure of the interface
\cite{selke}. It would appear that such analytic calculations took into
account only the `self-energy' diagrams of our analysis, not the
renormalisation of the branching vertex. As we show, the latter is
essential in deriving the existence of a critical value of $q$ from this
point of view. In addition, our RG analysis actually takes account of 
a potentially infinite number of nested branchings and self-energy
bubbles.

The layout of this paper is as follows. In the next section we define
more carefully the diagrammatic expansion we use to
model branching Potts interfaces in the SOS approximation, and argue that
it may be derived from a local Lagrangian field theory. Then we
discuss the renormalisation of this theory, calculating explicitly to
one loop order and deriving the results given above. 
Finally in Sec.~3 we discuss the random case
and make further comments on the problem. 

\section{Field theory of branching interfaces}
As discussed in the Introduction, the fact that configurations of branching 
Potts interfaces resemble Feynman diagrams for BARWs and DP does not imply 
that they are in direct correspondence. As is well known, a bare Feynman
propagator may be interpreted as the partition function 
$Z(t,x)$ for
sum of directed paths from $(0,0)$ to $(t,x)$ weighted by a factor 
$z_0^{\rm length}$: on the diagonal square
lattice, for example, such a partition function satisfies
\be
Z(t,x)=z_0\left(Z(t-1,x+1)+Z(t-1,x-1)\right)+\delta_{t,0}\delta_{x,0}
\ee
so that, in the infrared limit of interest, its Laplace-Fourier
transform has the form
\be 
G_0(s,k)=(s+D_0k^2+\sigma_0)^{-1}
\ee
where, in this example, $\sigma_0=2z_0-1$.
However, a sum over all such paths in a diagram such as Fig.~\ref{fig1}
would include configurations in which
paths corresponding to different propagators cross each other, and this is
unphysical in the case of interfaces. 

This may be taken into account by incorporating a repulsion between 
neighbouring walks, as follows.
Ignoring for the time being the $q$-dependent weight factors, 
the partition function for the interface pinned as described in the
introduction corresponds to the correlation function
\be
G(L,0)=\int{\cal D}\bar\phi{\cal D}\phi\,
\phi(L,0)\bar\phi(0,0)\,{e}^{-S}
\ee
where the action $S$ has the general form familiar from directed percolation
\cite{CS}
\be\label{S1}
S=\int dtdx[\bar\phi\partial_t\phi+D_0(\partial_x\bar\phi)
(\partial_x\phi)+\sigma_0\bar\phi\phi
-\ffrac12u_0(\bar\phi^2\phi+\bar\phi\phi^2)
+\ffrac14\lambda_0\bar\phi^2\phi^2]
\ee
In the absence of branching, $G=G_0$ as defined above. With branching,
we still expect that $G(L,0)\sim e^{-\sigma L}$ as $L\to\infty$, and so
we may interpret the renormalised `mass' $\sigma$ as the interfacial tension.
The term proportional to $\lambda_0>0$ provides an effective repulsion between
different pieces of the interface. Strictly we should take the limit of 
infinite $\lambda_0$, but, as with other self-repelling walk problems,
finite repulsion is expected to be in the same universality class. 
In fact, our explicit calculations will be for infinite coupling, which then
requires special treatment, described later.

\begin{figure}
\centerline{
\epsfxsize=4in
\epsfbox{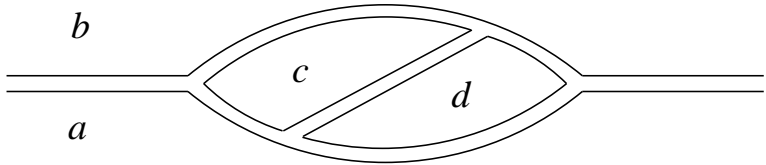}}
\caption{Fat graph corresponding to Fig.~1.}
\label{fig2}
\end{figure}

The diagrams generated by (\ref{S1}) do not have the correct weights to
represent branching Potts interfaces. While for a given diagram with a
small number of loops it is relatively easy to compute this weight, which is
simply the number of colourings of the internal loops consistent with
neighbouring regions being coloured differently, it is not clear how to give
a rule for doing this to all orders in such a way that the theory is
obviously multiplicatively renormalisable. This is because the colouring problem
appears to involve satisfying global constraints which are not simply
given by simple rules at the local level of the vertices. However, it is
possible to generalise (\ref{S1}) in such a way that these weights will
be generated from a local Lagrangian and thus, most likely, correspond to
a renormalisable field theory. Observe that each segment of interface
separates two different states of the Potts model, say $a$ and $b$. The
corresponding propagator may therefore be represented by a `fat' line
carrying labels $a$ and $b$, as shown in Fig.~\ref{fig2}. 
Such a fat 
line represents the propagator $\langle\phi_a^b(x,t)\bar\phi_a^b(0,0)\rangle$
of a \em matrix\em-valued field $\phi_a^b$. Since $a\not=b$, this transforms
according to a generically irreducible
representation of the permutation group of $q$ objects.
The branching vertices are now represented by interactions of the form
\be\label{S2}
-u_0\bar\phi_a^b\phi_a^c\phi_c^b+ {\rm h.c.},
\ee
and, in any given loop graph, the internal indices are summed subject to
all the constraints. Similarly, the repulsion between neighbouring
interfaces may be modelled by an interaction of the form
\be
\lambda_0\bar\phi^a_b\bar\phi^b_c\phi^a_b\phi^b_c
\ee

This construction still, however, includes unphysical non-planar graphs.
These may be removed by a well-known trick. Add additional $O(N)$ indices
$(i,j)$ to the fields, where $1\leq i,j\leq N$, and generalise
(\ref{S2}) to
\be\label{S3}
-u_0\bar\phi_{a,i}^{b,j}\phi_{a,i}^{c,k}\phi_{c,k}^{b,j}+ {\rm h.c.},
\ee
where the internal $O(N)$ indices are summed freely. On taking the limit
$N\to\infty$, $u_0\to0$, with $u_0^2N$ fixed, only the planar diagrams
survive. 

Having argued that the configurations of branching interfaces can be
viewed as the Feynman diagrams of a local Lagrangian
field theory, we shall in fact
make no further reference to this. Since we shall be concerned only with
low order calculations, the relevant diagrams and their weight factors will be
straightforward to write down by inspection.

\subsection{Fermionic representation and perturbative calculation}
It is possible to develop and renormalise the perturbative expansion of
$G$ simultaneously in powers of both the branching $u_0$ and
the repulsion $\lambda_0$. In the absence of branching there is a non-trivial
fixed point in transverse dimension $d<2$,
which corresponds to the limit of infinite
bare parameter $\lambda_0$. However, the cubic branching term becomes
marginal at $d=4$, just as in DP, and so it is strongly relevant for
$d<2$. It is of course possible to consider the truncated loop expansion
in both $\lambda_0$ and $u_0$, but there is no reason for it to give
sensible results for the physical case $d=1$. Indeed, the one-loop
calculations we have
performed for this case appear to lead to no relevant fixed point in $d=1$
describing the interface problem. 

Instead, we shall adopt a different strategy which is appropriate only in
the physical number of transverse dimensions $d=1$. In that case we may
observe that, in between branchings, the interfaces behave like the world
lines of non-relativistic bosons with infinitely strong hard core
repulsion. In one dimension, these are equivalent to free fermions. The
interfaces thus propagate as fermions in between branching events.
However, since no more than one particle may then occupy the same site, it
now becomes necessary to smear the branching process
so that particles are shifted to
neighbouring sites of the lattice. Of course this should not affect the
universal behaviour.

\begin{figure}
\centerline{
\epsfxsize=3in
\epsfbox{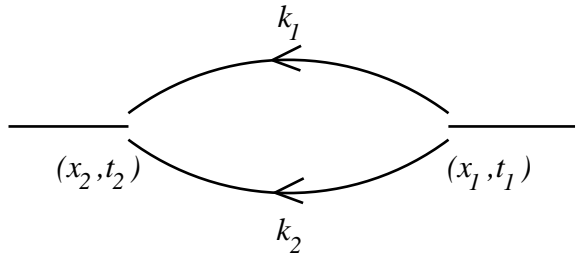}}
\caption{Point-split version of the one-loop contribution to 
$\Gamma^{(1,1)}$.}
\label{fig3}
\end{figure}
Consider, for example, the simple bubble diagram shown in 
Fig.~\ref{fig3}, evaluated in real space initially. The pair of walks in the 
loop begin and end at $(t_1,x_1)$ and $(t_2,x_2)$ respectively, but we
separate them so these become $(t_1,x_1\pm a)$ and 
$(t_2,x_2\pm a)$, where $a$ is of the order of the lattice spacing.
The propagator for such a pair of walks which avoid each other is given
by the method of images as
\begin{eqnarray}
&&G_0(t_1,x_1+a;t_2,x_2+a)G_0(t_1,x_1-a;t_2,x_2-a)\nonumber \\
&&-G_0(t_1,x_1+a;t_2,x_2-a)G_0(t_1,x_1-a;t_2,x_2+a),
\end{eqnarray}
or, in terms of Fourier transforms
\be
\int {dk_1dk_2\over s+D_0(k_1^2+k_2^2)+2\sigma_0}
\left(e^{ik_1(x_1-x_2)}e^{ik_2(x_1-x_2)}-e^{ik_1(x_1-x_2+2a)}
e^{ik_2(x_1-x_2-2a)}\right)
\ee
Expanding the last factor to lowest non-vanishing order then gives 
$\frac12(2a)^2(k_1-k_2)^2$. Each vertex thus acquires a wave number
dependence $\propto(k_1-k_2)$. The factor $\frac12$ is the usual symmetry
factor for this diagram. We may absorb the factors of $2a$ into the
branching rate $u_0$ to give a new effective coupling constant
$\tilde u_0\equiv 2au_0$. Notice this will have a different canonical
dimension. The calculation of Fig.~\ref{fig3} is therefore equivalent to
that in a theory with Grassmann fields $c(x,t)$, $\bar c(x,t)$,
with the same propagator but with interactions
\be\label{eq9}
\ffrac12\tilde u_0(\bar c\partial_x\bar c\,c+\bar c\,c\partial_xc)
\ee
Note that this is the lowest order non-vanishing cubic interaction consistent
with the rules $c^2={\bar c}^2=0$.

\begin{figure}
\centerline{
\epsfxsize=3.5in
\epsfbox{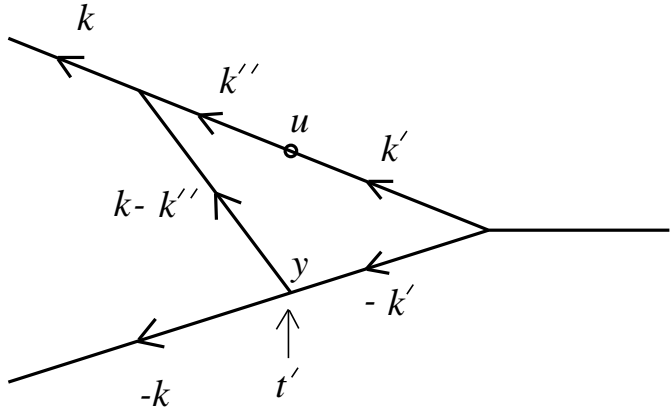}}
\caption{One-loop renormalisation of the coupling constant. In this
diagram, the physical region is restricted to $u>y$.}
\label{fig4}
\end{figure}

In general, this fermionic correspondence extends to higher order
diagrams. There is one important complication however. The
interfaces behave as free fermions only in between branching or
coagulation events, wherever these may be, which may involve other
particles. An example is the one-loop renormalisation of the vertex,
shown in Fig.~\ref{fig4}. 
In this case the interface passing through the
point $u$ in the figure must avoid the branching at point $y$, which
occurs at the same `time' $t'$. While the fermionic formulation ensures that
the interfaces immediately to the right and left of $t'$ avoid each other,
if we were to take the fermionic lagrangian literally we would integrate
over all values of the intermediate coordinate
$u$. This would include the region $u<y$ which is
unphysical and where, in fact, the amplitude has the wrong sign.
Thus, in using the fermionic formulation, we must be careful to
integrate over only the physical part of the phase space, where the
`particles' retain their original ordering. This is in
contrast to what would be done in a fully fermionic quantum field theory,
where it would be necessary to integrate over all of phase space so as
to respect the Fermi statistics of the wave functions.

The restriction to $u>y$ makes the calculation of the Feynman integral
considerably more complicated. However, we shall argue that as
long as we are interested only in the renormalisation of the infrared
limit of 
the vertex, we need simply to multiply the unrestricted diagram by a factor
$\frac12$. Similar rules are expected to hold for more complicated
graphs.

We are now in a position to summarise the Feynman rules:
\begin{enumerate}
\item Propagator $(s+D_0k^2+\sigma_0)^{-1}$;
\item Vertices $\pm i(k_1-k_2)\tilde u_0$, where $k_1$ and $k_2$ are the
wave-numbers of the pair of outgoing (incoming) fermions;
\item Usual symmetry factors;
\item $q$-dependent weights corresponding to the number of colourings of
the diagram. There will also be factors of $(-1)$ from fermion Wick
contractions. These in fact cancel minus signs from the vertices. The
overall sign of a diagram is, in any case, always positive for
sufficiently large positive $q$.
\item Factors due to the restriction of phase space, as discussed above.
\end{enumerate}

\subsection{One-loop renormalisation}
In the usual way, we may define the one-particle irreducible vertex
functions $\Gamma^{(1,1)}$ and $\Gamma^{(2,1)}$
for this theory and compute the renormalisation of the fields, the
diffusion constant and the coupling constant \cite{DRG}. We adopt the scheme of
renormalising in the massive theory ($\sigma>0$) at zero external
frequency and wave number, which makes the integrals somewhat simpler
\cite{ID}.
From Fig.~\ref{fig3} we have
\begin{eqnarray*}
\lefteqn{\Gamma^{(1,1)}(s,k)=s+D_0k^2+\sigma_0}\nonumber\\
& &{}-\ffrac12(q-2){\tilde u}^2_0\int{dk'\over2\pi}
{(2k')^2\over s+D_0(k'+\ffrac12k)^2+D_0(k'-\ffrac12k)^2+2\sigma_0}
+O({\tilde u}^4_0)
\end{eqnarray*}
Imposing the normalisation condition
\be
(\partial/\partial s)\Gamma^{(1,1)}_R(0,0)=1
\ee
with $\Gamma^{(1,1)}_R=Z_c\Gamma^{(1,1)}$, yields the field
renormalisation
\be 
Z_c=1-\ffrac12(q-2){\tilde u}^2_0\int{dk'\over2\pi}
{4{k'}^2\over(2D_0{k'}^2+2\sigma_0)^2}
=1-\ffrac18(q-2)g_0^2+O(g_0^4),
\ee
where $g_0^2={\tilde u}^2_0/D_0^{3/2}\sigma_0^{1/2}$.
Similarly, we find for the renormalised diffusion constant 
\be
D_R\equiv{\partial\over\partial k^2}\Gamma^{(1,1)}_R(0,0)
=Z_cD_0\left(1+\ffrac1{16}(q-2)g_0^2+O(g_0^4)\right)
\ee

Next, we have the vertex renormalisation, given at one loop by two
diagrams like Fig.~\ref{fig4}. 
The calculation of this diagram is complicated by the restriction $u>y$.
For this reason it is not true that the wave number flowing through the
point $u$ is simply conserved. Instead, the integration over $u$ yields
\be
\int^{\infty}_y{e}^{-ik'u+ik''u-\epsilon u}du
={-i{e}^{-ik'y+ik''y}\over k'-k''-i\epsilon},
\ee
where we have inserted the factor ${e}^{-\epsilon u}$, with 
$\epsilon\to0+$, to ensure convergence. Now use the identity
$(x-i\epsilon)^{-1}=P(1/x)+i\pi\delta(x)$, and observe that the whole
diagram is invariant under simultaneous complex conjugation and sending
$k\to-k$. The above decomposition will therefore lead to two terms with
opposite parity under $k\to-k$. The second term is precisely one half of
what would be obtained if we had ignored the restriction $u>y$, and, as
we see explicitly below, its has the form $ik$ times a real
function of $k^2$. By contrast, the first term must therefore yield
a real function of $k^2$ only. Since we are interested in the
contribution $\propto ik$ in
$\Gamma^{(2,1)}(k,-k,0)$ in the infrared limit $k\to0$,
it follows that we need retain only the second term, which is one half
of what we would have found imposing wave number conservation $k'=k''$.

The result is therefore
\be
\Gamma^{(2,1)}=2ik\tilde u_0
+\ffrac122(q-3){\tilde u}^3_0
\int{dk'\over2\pi}
{2ik'\,i(k-k')\,i(2k-k')\over(2D_0{k'}^2+2\sigma_0)
(D_0{k'}^2+D_0(k-k')^2+2\sigma_0)}
\ee
where the other important feature to note is the factor of $q-3$ which comes
from the fact that there are three different colours on the boundary,
and the internal colour must differ from each of these.
In this integral we
need take only the term $O(k)$ as $k\to0$. After some straightforward
quadratures we find
\be
\Gamma^{(2,1)}=2ik\tilde u_0\left(1+\ffrac7{32}(q-3)g_0^2+O(g_0^4)\right)
+O(k^3)
\ee

From these expressions we may find the dimensionless renormalised
coupling 
\be
g\equiv{(2ik)^{-1}\Gamma^{(2,1)}Z_c^{3/2}\over\sigma^{1/4}D_R^{3/4}}
={{\tilde u}_0\over\sigma^{1/4}D_0^{3/4}}
\left[1+\left(\ffrac7{32}(q-3)-\ffrac9{64}(q-2)\right)g_0^2+\cdots
\right]
\ee
Notice that $g_0$ depends on the bare mass $\sigma_0$, but, to the order
indicated, we may replace this by $\sigma$.
We then finally have the beta-function
\be
\beta(g)\equiv\sigma{\partial g\over\partial\sigma}\Big|_{\tilde
u_0}
=-\ffrac14g-\ffrac12\left(\ffrac7{32}(q-3)-\ffrac9{64}(q-2)\right)g^3
+\cdots
\ee

This is the main result (\ref{eq1}) quoted in the introduction.
We have left the second term in this form to illustrate how the critical
value $q_c\approx q_0
=\frac{24}5$ (in this approximation) emerges from the vertex
renormalisation, proportional to $q-3$, and that of the propagator,
proportional to $q-2$. In particular, one may see how the latter
increases $q_c$ above the naively expected value of three.
This also illustrates why the hierarchical lattices of Ref.~\cite{Kardar}
give a different result, for in these cases there are only ever factors
of $q-2$, and no vertex corrections.

\subsection{Calculation of the Widom exponent}
As discussed in the introduction, the renormalised interfacial tension
$\sigma$ is supposed to vanish close to a continous transition
as $(\sigma_0-\sigma_{0c})^\mu$. In the language of directed
percolation, $\mu$ may therefore be identified with the exponent
$\nu_\parallel$ (or, in the language of dynamic critical behaviour,
$z\nu$.) The method of estimating its value using the field-theoretic
renormalisation group is standard \cite{DRG}. Define the vertex $\Gamma^{(1,1)}$
with an insertion of $\bar cc$ as
\begin{eqnarray}
\Gamma^{(1,1;1)}\equiv(\partial/\partial\sigma_0)\Gamma^{(1,1)}(0,0)
&=&1+(q-2){\tilde u}_0^2\int{dk'\over2\pi}
{(2k')^2\over(2D_0{k'}^2+2\sigma_0)^2}+\cdots\nonumber\\
&=&1+\ffrac14(q-2)g_0^2+\cdots
\end{eqnarray}
In turn, this gives the combination of renormalisation constants
$Z_c^{-1}Z_{\bar cc}$, if we normalise so that $\Gamma^{(1,1;1)}_R=1$
at zero frequency and wave number.
Thus we find that
\be
\gamma_{\bar cc}\equiv\sigma{\partial\ln Z_{\bar cc}\over\partial
\sigma}=
-\ffrac1{16}(q-2)g^2+\cdots
\ee
In the standard way, at the fixed point $\gamma_{\bar
cc}^*=1-\nu_\parallel^{-1}$, which yields, in this one-loop
approximation
\be\label{eq20}
\mu=\nu_\parallel\approx{24-5q\over20-3q}
\ee
Obviously this is only a rather crude estimate and it is difficult
to gauge its accuracy. However it does exhibit the correct trend
as $q$ varies close to 2, and in fact for $q=3$ gives the remarkably
good estimate $\mu\approx\frac9{11}$ to be compared with the exact value
$\frac56$.

In the context of critical dynamics or DP it is also
possible to define two other independent
critical exponents, namely the dynamical exponent $z$ and $\eta_\parallel$, 
defined by the scaling law
\be
G(t,k)\sim t^{-\eta_\parallel}\,F(tk^z)
\ee
for the Fourier transform with respect to $x$ of $G(x,t)$, evaluated at
criticality. For simple non-branching directed paths we have $z=2$
and $\eta_\parallel=0$. The latter result agrees with that expected for
an Ising interface whose ends are pinned at the points $(0,0)$ and
$(L,x)$, afer the partition function, which corresponds to $G(L,x)$, is
integrated over $x$. This is because the pinning sites correspond to
insertions of disorder operators at the boundary points $(0,0)$ and
$(x,L)$: these are expected to scale as $L^{-2x_s}$, where $x_s=\frac12$ 
is the
boundary scaling dimension of such an operator, equal by duality to the
boundary dimension of the magnetisation. The additional integration over
$x$ then provides another factor of $L$, giving
$\eta_\parallel=2x_s-1=0$. On the other hand, the result $z=2$ may not
simply be compared with bulk Ising exponents, since it refers to the
SOS picture of the interface, which neglects overhangs and bubbles of
the opposite phase and therefore cannot be directly compared with, for
example, the width of the magnetisation profile which is expected to
scale as $L^1$ at the critical point.

Within the field-theoretic formulation of the interface for $q>2$, these
two exponents are related, in the standard way, to the fixed point
values of the renormalisation group functions 
$\gamma_c\equiv\sigma(\partial/\partial\sigma)\ln Z_c$, and 
$\gamma_D\equiv2\sigma(\partial/\partial\sigma)\ln Z_D$, where
$D_R=Z_DD_0$. Explicitly, $\eta_\parallel=\gamma_c^*$ and
$z=2+\gamma_D^*$.
In the one-loop approximation, this gives
\begin{eqnarray}
\eta_\parallel&\approx&\ffrac1{16}(q-2)g^2\approx\nu_\parallel^{-1}-1
\label{eq23}\\
z&\approx&2+\ffrac1{16}(q-2)g^2\approx\nu_\parallel^{-1}+1\label{eq24}
\end{eqnarray}

For $q=3$, for example, this leads to the estimate
$\eta_\parallel\approx\frac29$, to be compared with the exact value
$2x_s-1=\frac13$ \cite{CSURF}. 
For $z$ we find $\approx\frac{20}9$, consistent with
this measure of the interfacial width becoming larger as the interface
becomes more branched. 

\subsection{The directed percolation limit} 
As discussed in the introduction, the configurations of a branching
Potts interface resemble the diagrams of the field theory of directed
percolation (DP) \cite{CS}, but, in general, the weights are different, since in
DP these are simply given by a factor $(-1)$ for each closed loop. This
weighting may be recovered, however, by taking the formal limit
$q\to-\infty$ with $\tilde u_0^2(-q)$ fixed. if we take this limit in
Eqs.~(\ref{eq20},\ref{eq23},\ref{eq24})
we find the estimates $\nu_\parallel\approx1.67$,
$z\approx1.6$ and $\eta_\parallel\approx-0.4$ (the latter exponent gives
the rate of growth of the average number of infected sites $t^{-\eta_\parallel}$
from a single seed). As discussed in the introduction, these agree
remarkably well with currently accepted values from enumerations and
Monte Carlo methods. This success at the one-loop approximation may be
attributed to the inclusion of the infinitely strong repulsion between
neighbouring walks. As discussed above, this implies that the effective
fermionic interaction vanishes at zero wave number, and as a consequence
the canonical dimension of the effective coupling is reduced from
$[u_0]=k^{3/2}$ to $[\tilde u_0]=k^{1/2}$. Equivalently, one may say
that the upper critical dimension of the field theory is reduced from
$d=4$ transverse dimensions to $d=2$ (although, strictly speaking, the
fermionic theory makes sense only for $d=1$). 
Thus we expect the loop expansion
to give more accurate results when truncated at a low level.
It would be interesting to extend this analysis to two loops.

However, the fact that the theory when expanded about infinitely large
quartic coupling $\lambda_0$ gives exponents which appear to agree with
those obtained from the theory with zero bare quartic coupling is
non-trivial, and provides
further evidence for the robustness of the DP universality class.

\section{Further remarks} 

\subsection{Effect of randomness} 
As discussed in the introduction, we were originally motivated to study
this problem by the paper of Kardar et al.~\cite{Kardar} on the effect of
bond disorder on the branching Potts interface. Their model, on certain
types of hierarchical lattice, is unphysical in that exhibits a
continuous transition for all $q$ in the pure case, but it does have one
important feature in common with Potts models on regular 2d lattices:
namely that the randomness is relevant for $q>2$, and marginally irrelevant
for the Ising case $q=2$. This feature persists to the interfacial model
since it is exact on the hierarchical lattice. 
This enabled Kardar et al.~\cite{Kardar} to draw 
exact conclusions for their model, in particular
that the branching is marginally irrelevant. Whether these
conclusions also hold for the random Potts models on regular lattices is
of course debatable, and in the light of recent evidence \cite{Picco,JJ}
probably incorrect.

It would be very useful if randomness could be incorporated into a
more realistic yet solvable interface model. Unfortunately the SOS
approximation, while it captures most of the features of the pure model,
fails to do so in the random case. This may already be seen for $q=2$.
In that case, the random bonds act as randomly positioned pinning
centres for the interface. This is equivalent to the well-studied
problem of a directed polymer in a random medium, which may be mapped
onto the KPZ equation \cite{KPZ}. Unlike the case of the bulk random bond Ising
model, the randomness in this SOS model is strongly relevant, changing,
in particular, the dynamic exponent $z$ from two to the KPZ value of
$\frac32$. This implies that the SOS model is not a useful one in which
to study the effects of randomness. Indeed, it shows that the contributions
of overhangs and bubbles of the opposite phase do not simply cancel in
the random case, at least near bulk criticality.

Nevertheless, it is interesting to investigate whether branching of such
randomly pinned directed polymers is relevant. A simple scaling argument
in fact indicates that it should be marginal, as in the case studied by
Kardar et al \cite{Kardar}. If we use the fermionic lagrangian of
(\ref{eq9}) we may extract the following dimensions in terms of transverse
wave number $k$ and `frequency' $\omega$: $[\bar cc]=k$, implying that
$[\bar c]=[c]=k^{1/2}$. Thus $[{\tilde u}_0]\cdot k\cdot k^{3/2}=\omega k
=k^{1+z}$. Using the KPZ value $z=\frac32$ we see that ${\tilde u}_0$ is
dimensionless, and therefore marginal. Unfortunately it is much more
difficult to determine whether or not it is marginally irrelevant.
It would be interesting to understand whether the marginality of
branching is a stable property of pinned directed interfaces, or whether
the similarity between the hierarchical and euclidean 2d cases is
coincidental.

\subsection{The effect of the disordered phase}
Although the one-loop approximation to the renormalisation group
equations for the SOS branching Potts interface captures much of the
physics of the pure Potts model, even quantitatively, there is one 
important feature lacking, which is the description of the behaviour as
$q\to q_c$. If we integrate (\ref{eq1}) for $q>q_c$ to calculate the 
renormalised interfacial tension
at which $g$ diverges (which we interpreted as the first-order
transition) we find 
$\sigma\sim\sigma_0{e}^{-1/2b(q-q_c)}$. This is to be compared with
the exactly known 
behaviour $\sim{e}^{-{\rm const}/(q-q_c)^{1/2}}$ \cite{Baxter}.
Similarly, our model suggests that, as $q\to q_c-$, $\mu\to0$, while
in fact it retains a finite value at $q=q_c=4$, jumping discontinuously
to an (effective) value of zero for $q>4$. 

These two features, which are missing in our approach, are qualitatively
accounted for in the picture of the fixed point structure which emerges
from approximate real space RG calculations which incorporate annealed
vacancies as additional degrees of freedom \cite{Nienhuis}. These suggest
that, for $q<q_c$, there are two fixed points, corresponding respectively
to critical and
tricritical behaviour, which merge in a parabolic fashion at $q=q_c$,
thus giving rise to a marginal operator at this point. In fact, this
qualitative picture has been shown to lead to all sorts of generalised
scaling relations near $q=4$ which have been verified by exact results
and by Monte Carlo simulations \cite{CNS,Sokal}.

It is clear what is missing in our model for $q\geq q_c$. In this case,
at the bulk critical point, there is coexistence not only between the
$q$ ordered phases, but also with the disordered phase. Thus, in the
interfacial model, we should take into account not only interfaces
between different ordered phases, but also those between any ordered
phase and the disordered phase. Both types of renormalised
interfacial tension are
expected to vanish at bulk criticality. Indeed, it has been found both
numerically and analytically \cite{wetting} that the disordered phase
wets the interface between two ordered phases at $T=T_c$ for $q>q_c$.
The resulting SOS model has,
potentially, a very rich structure. Each type of interface will have its
own bare `mass' or interfacial tension, and, in principle, its own
diffusion constant. In addition, there are now three different types of
branching process which reflect themselves in three \em a priori \em
independent couplings to be renormalised. It remains to be seen whether
the known behaviour for $q>q_c$ may be fitted into this picture, but,
given that vacancies may be regarded as microscopic regions of
disordered phase, there is every expectation that this should be the
case. It would also be interesting to apply the ideas of this paper
to other more
complicated models (e.g. the Ashkin-Teller model) with more than one
type of interface.

I woould like to thank S.~Dietrich and W.~Selke
for directing me towards the earlier
literature on Potts interfaces.
This work was supported in part by EPSRC Grant GR/J78327.

\end{document}